\documentclass[12pt,a4paper]{article}
\usepackage{epsfig,multicol}
\usepackage{amsmath,amssymb}
\usepackage{bbm}
\usepackage[latin1]{inputenc}
\usepackage{fancyhdr}
\usepackage{cite}

\numberwithin{equation}{section}

\def\bpm{b^{+-}}
\def\bpp{b^{++}}

\def\Dmm{\mathrm{D}^{--}}
\def\Dpp{\mathrm{D}^{++}}

\def\nablapp{\nabla^{++}}

\def\ppp{\partial^{++}}

\def\thetabm{\bar{\theta}^-{}}
\def\thetabp{\bar{\theta}^+{}}
\def\thetam{\theta^-{}}
\def\thetap{\theta^+{}}
\def\Vmm{V^{--}}

\def\vmm{v^{--}}
\def\vpp{v^{++}}

\def\iim{\mathrm{i}}
\def\V{V^{++}_{\text{WZ}}}

\def\ve{\varepsilon}
\def\phib{\bar{\phi}}
\def\Psib{\bar{\Psi}}

\def\thetab{\bar{\theta}}

\def\alphad{{\dot{\alpha}}}
\def\betad{{\dot{\beta}}}

\def\Ac{{\cal A}}
\def\Fc{{\cal F}}

\def\Nc{{\cal N}}
\def\Wc{{\cal W}}

\def\iim{\mathrm{i}}
\def\Oh{\mathrm{O}}

\title{Non-singlet Q-deformed $\Nc=(1,0)$ and \\ $\Nc=(1,1/2)$  U(1) actions}

\author{A. De Castro and L. Quevedo}

\date\empty
\begin{document}

\maketitle
\setcounter{page}{1}

\begin{center}

Institut für Theoretische Physik, Universität Hannover, \\
Appelstrasse 2, 30167 Hannover, Germany. \\
{\tt castro, quevedo@itp.uni-hannover.de}
\end{center}
\vspace{5mm}
\thispagestyle{fancy}
\fancyhead[RO]{ITP-UH 11/06 \\ hep-th/0605187}

\begin{abstract}
\noindent{}In this paper we construct $\Nc=(1,0)$ and $\Nc=(1,1/2)$
non-singlet Q-deformed supersymmetric U$(1)$ actions in components. We
obtain an exact expression for the enhanced supersymmetry action by
turning off particular degrees of freedom of the deformation tensor. We
analyze the behavior of the action upon restoring weekly some of the
deformation parameters, obtaining a non trivial interaction term between
a scalar and the gauge field, breaking the supersymmetry down to
\mbox{$\Nc=(1,0)$}.  Additionally, we present the corresponding set of unbroken
supersymmetry transformations. We work in harmonic superspace in four
Euclidean dimensions.
\end{abstract}



\section{Introduction}
\label{generalities}
Being deformations of field theories and supersymmetry both old ideas
deeply analyzed and developed through many decades, it is a natural step
to think of extending the Weyl-Moyal product to a deformed algebra
of superfields involving the Graßmann sector, thus leading to
\emph{non-(anti)commutativity}.  Recently it has been found that strings
in certain backgrounds are related to such deformations of superspace,
see for example \cite{Seiberg:2003yz, Berkovits:2003kj, deBoer:2003dn,
Ooguri:2003qp, Ferrara:2004zv}. This has stimulated the study of
particular supersymmetric gauge theory deformations, implemented through
an associative algebra of superfields whose Moyal product is realized
as a function of a bilinear nilpotent Poisson operator. For this reason
this formulations are also called {\it Nilpotent deformations}.  A very
interesting feature of such non-(anti)commutative theories is the
natural emergence of interactions not present in the corresponding
undeformed scenarios. For instance, in our case we will see the
apparition of Yukawa-like interactions. The progress made towards the
understanding on the renormalizability of non-anticommutative field
theories \cite{Grisaru:2005we, Britto:2003aj, Buchbinder:2005sh} is also
very motivating. 

Nilpotent deformations in extended supersymmetric field theories were
first analyzed in superspace \cite{Ferrara:2003xy, Klemm:2001yu} and
later on in harmonic superspace \cite{Ivanov:2004qz, Ivanov:2003te}. In
this paper we work in Euclidean harmonic superspace in four dimensions
\cite{Galperin:2001uw}, where $(\theta^{\alpha}_i)^* \neq
\thetab^{\alphad}_i$. In general, nilpotent deformations are introduced
via Weyl-Moyal product with a bilinear Poisson operator which is
constructed either in terms of the supercharges, or in terms of the
spinor covariant derivatives \cite{Ferrara1, Klemm:2001yu,
Ferrara:2003xy}, leading to \emph{Q-} and \emph{D-deformations},
respectively.  Like in non-commutative field theories, even when there
is no unique non-anticommutative generalization of a given
supersymmetric theory, a selection scheme can be found based on
different physical reasons like symmetry preservation or its relation to
string theory.  Since Q-deformations are directly implied by string
theory, it seems tempting to continue studying their physical
properties, postponing the identification of the specific string
backgrounds from which the resulting theories originate.  Therefore we
concentrate our analysis in Q-deformations. In this paper we construct
the exact Q-deformed supersymmetric $\Nc=(1,1/2)$ action by dropping
consistently some components of the deformation matrix.  Afterwards,
while weakly restoring some degrees of freedom of the deformation
parameters, we break the supersymmetry down to $\Nc=(1,0)$ obtaining a
second order action.  As we will see in \S \ref{actions} we can also
choose this variables to control a certain potentials appearing from the
deformation of the $\Nc=(1,1)$ which can not be disentangled by
redefinitions of the components fields. We calculate the corresponding
expressions for the full set of non-singlet Q-deformed supersymmetry
transformations, together with the Seiberg-Witten-like map which sets a
frame where actions are gauge invariant under the canonical undeformed
transformations. Though all our actions have partially broken
supersymmetry, it has been shown they preserve the so-called twist
supersymmetry \cite{Zupnik:2006qa} by construction.
\section{Non-singlet Q-deformations and supersymmetry\\ breaking}
In general terms the Poisson operator is written as\footnote{In the
sequel, we use all conventions in \cite{DeCastro:2005vb}. Note that all
operators are left derivations unless explicitly stated. The only right
derivation appears on the Poisson operator.}
\begin{equation}
P = -\overleftarrow{Q}^i_\alpha C^{\alpha \beta}_{ik}
\overrightarrow{Q}^k_\beta\,,
\label{Poisson}
\end{equation}
where Greek letters represent Euclidean space-time indexes $\alpha,\beta=1,2$
and $\alphad,\betad=\dot{1}, \dot{2}$, whereas Latin indexes stands for SU(2)
automorphisms $i,j=1,2$. Both
sorts of indices are raised and lowered with the SU(2) metric
$\ve_{\alpha\beta}, \ve_{\dot\alpha\dot\beta}, \ve_{ik}$,
where $\ve_{12}=1$. The Moyal product of two superfields is then defined by
\begin{equation}
A\star B=A e^P B=AB+APB+\frac12 AP^2B+\frac16 AP^3B+\frac1{24} AP^4B, \quad P^5=0\,. \label{MP}
\end{equation}
in order to preserve the associativity of the Moyal product,
\eqref{Poisson} is chosen to include only undotted supercharges. From
\eqref{MP} we see that the nilpotent nature of the Poisson operator
\eqref{Poisson} advantageously makes the Moyal product polynomial,
producing local deformed theories.  Q-deformations, in contrast to
D-deformations\footnote{See for example \cite{Ketov:2004yt}}, break
supersymmetry, but preserve chirality and, in the ${\cal N}=(1,1)$
harmonic superspace case, also Graßmann harmonic analyticity and the
harmonic conditions $\mathrm{D}^{\pm\pm}A=0$, which  are preserved in virtue of the
properties \cite{Ivanov:2003te}
\begin{equation}
\left[D^{\pm}_\alpha , P\right] = 0\,, \qquad
\left[\bar{D}^{\pm}_{\alphad} , P\right] = 0\,, \qquad
\left[\mathrm{D}^{\pm\pm} , P\right] = 0,
\end{equation}
A proper definition of the anticommutators involving the bilinear
operator $P$ is given by
\begin{equation}
A[\epsilon\cdot G,P]B\equiv-C^{\alpha\beta}_{ij}\left( 
[\epsilon\cdot G,\partial_\alpha^i]A\,\partial_\beta^jB
+\partial_\alpha^iA\,[\epsilon\cdot G,\partial_\beta^j]B\right).
\label{Comm}
\end{equation}
When $G$ is the generator of a symmetry $\delta_\epsilon A=-\epsilon^aG_aA$, the
commutator above measures to what extent the Moyal product breakes the Leibniz
rule for its transformation laws
\begin{equation}
\delta_\epsilon (A\star B) \neq\delta_\epsilon A\star B+A\star\delta_\epsilon B.
\end{equation}
The deformation parameters ${C}^{\alpha\beta}_{ij}={C}^{\beta\alpha}_{ji}$ form
a constant tensor which can be split in the following way \cite{Ferrara:2003xy,
Ivanov:2003te}
\begin{equation}
{C}^{\alpha\beta}_{ij} = I  \ve^{\alpha \beta}\ve_{ik}
 +\hat{C}^{\alpha \beta}_{ik}\,. \label{Generic}
\end{equation}
Q-deformations induced by the first term are called \emph{singlet} or
\emph{QS-deformations}, whereas those associated with the second term
can naturally be named \emph{non-singlet} or \emph{QNS-deformations}.
The singlet term in \eqref{Generic} is $Spin(4)\times$SU(2)-preserving
while the non-singlet term involves a SU(2)$_\text{L}\times$SU(2)
constant tensor which is symmetric under independent permutations of
Latin and Greek indices.  $\hat{C}^{\alpha \beta}_{ik} $ in general
breaks the space-time and R-symmetry groups $Spin(4)\times$
O(1,1)$\times$ SU(2) $\equiv$ SU(2)$_{\text{L}}\times$
SU(2)$_{\text{R}}\times$ O(1,1)$\times$ SU(2) down to SU(2)$_\text{R}$.
Nevertheless, choosing a particular factorizable form
\begin{equation}\label{hatcdecomp}
\hat{C}^{\alpha\beta}_{ij}=c^{(\alpha\beta)}b_{(ij)}\,.
\end{equation}
we are able to recover part of the symmetry group leaving
U(1)$_{\text{L}}\times$ SU(2)$_{\text{R}}\times$ U(1)
unbroken. We will see that the bosonic sector of our resulting actions are
manifestly invariant under the complete space-time and R-symmetry group.
It is clear that, with this matrix decomposition, we discard three
degrees of freedom among the nine parameters of the generic non-singlet
tensor, leading to the maximal symmetry preserving selection. Observing
the structure of the Moyal product in Q-deformations, it is not hard to
realize that theories constructed in this frame will have at least $1/4$
supersymmetries lost. A simple way to see this is by looking at the
only nontrivial commutator \eqref{Comm} for supersymmetry charges, namely
\begin{equation}\label{qbreak}
A[\bar\epsilon\cdot\bar{Q},P]B=2\iim 
\left( I \ve^{\alpha \beta}\ve_{ij}+\hat{C}^{\alpha\beta}_{ij}\right)
\bar{\epsilon}^{\alphad i}
\left( \partial_{\alpha\alphad}A\,\partial^j_{\beta}B
-\partial^j_{\beta}A\,\partial_{\alpha\alphad}B \right).
\end{equation}
Here we can appreciate that for a generic tensor $C^{\alpha\beta}_{ij}$
as well as for any value of $I$ in a singlet deformation, ${\cal
N}=(1,1)$ supersymmetry is broken to ${\cal N}=(1,0)$
\cite{Ferrara:2004zv}. Only for particular purely non-singlet parameters
we are able to enhance the supersymmetry to $\Nc=(1,1/2)$
\cite{Ivanov:2004hc, Araki:2005pc}. For example, from 
\begin{equation}\label{c11}
\hat{C}^{\alpha\beta}_{11}\neq0\,,\qquad 
\hat{C}^{\alpha\beta}_{12}=\hat{C}^{\alpha\beta}_{22}=0
\end{equation}
and \eqref{qbreak}, the commutation of
$\bar{\epsilon}^{\alphad2}\bar{Q}_{\alphad2}$ with $P$ obviously follows.
Therefore, implementing \eqref{c11}, the supersymmetry is broken down to
${\cal N}=(1,1/2)$ recovering the $1/4$ fraction generated by
$\bar{Q}_{2\alphad}$. The exact expressions for non-singlet gauge and
supersymmetric transformations for the U(1) vector multiplet with
\eqref{c11} were first constructed in \cite{Araki:2005pc}, where the
authors also constructed the ${\cal N}=(1,1/2)$ invariant action in
components to first order in the deformation $C$. In
\cite{DeCastro:2005vb} we constructed the bosonic action using the
maximally space-time and R-symmetry preserving  parameters in
\eqref{hatcdecomp} for the generic case and with $b_{ij}$ restricted to
\begin{equation}\label{brestricted}
b_{11}\neq 0\qquad b_{12}=b_{22}=0.
\end{equation}
which is easily seen to be equivalent to the general solution for
vanishing determinant $b^2=\ve^{ik}\ve^{jl}b_{ij}b_{kl}=0$. In this case
$b_{ij}$ has rank 1 and admits a tensor product decomposition
$b_{ij}=b_ib_j$. By means of an appropriate SU(2) rotation one can pick
\eqref{brestricted} without loss of generality.  
\section{Non-singlet ${\cal N}=(1,0)$ and  ${\cal N}=(1,1/2)$ Q-deformed
actions}
\label{actions}
We start from the $\Nc=(1,1)$ Abelian gauge multiplet in four
dimensional Euclidean harmonic superspace. As we pointed out in the
introduction, the corresponding non-singlet Q-deformed models have some
fractions of the original supersymmetry broken. Though valuable effort
has been done obtaining the deformed components action in powers on the
full set of deformation parameters \cite{Araki:2004de, Araki:2004cv,
Araki:2004mq, Araki:2005pc}, it is clear that obtaining  exact
expresions for the deformed action in the general case is a very
difficult task, even using the matrix decomposition \eqref{hatcdecomp}.
For the pure bosonic case \cite{DeCastro:2005vb} we found a closed from
of the action using \eqref{hatcdecomp} and moreover, we were able
to redefine the fields in such a way that the Lagrangian took a particular
factorized form $\cosh^2(2\phib\sqrt{b^2 c^2})L_0$ where $L_0$ is the
free undeformed Lagrangian. In the present full supersymmetric case it
seems to be not an easy labor to accomplish that kind of simplicity.
Nevertheless, it is worthy to analyze different possibilities of the
non-singlet deformed action coming from selecting particular structures
of the deformation tensor $b_{ij}$.  For example, we can interpret
$b_{ij}$ as the set of supersymmetry breaking tuning parameters, i.e.
specific selections of $b_{ij}$ distinguish between different theories
with ${\cal N}=(1,0)$,  ${\cal N}=(1,1/2)$ and ${\cal N}=(1/2,1/2)$
supersymmetry, some of them with very simple Lagrangians. 

For our purposes, the most appropriate QNS-deformed $\Nc=(1,1)$ U(1)
gauge theory action in harmonic superspace \cite{Galperin:2001uw}, is
written in terms of the covariant superfield strength
\cite{Ferrara:2004zv}
\begin{equation}\label{curvaturaw}
    \Wc=-\frac14(\bar{D}^+)^2 \Vmm
\end{equation}
for which the action takes the form
\begin{equation}
    S=\frac14\int d^4x_L\,d^4\theta\,du\,\Wc\star\Wc=
    \frac14\int d^4x\,d^4\theta\,du\,\Wc^2\,.
\end{equation}
Note that $\Vmm$ is non analytic. A general expansion in
components reads
\begin{align}
\label{Vmm}
\nonumber\Vmm=\vmm+\thetab^+_\alphad v^{(-3)\;\alphad}
&+\thetab^-_\alphad v^{-\;\alphad}
+(\thetabm)^2 \Ac+(\thetabp\thetabm)\varphi^{--}
+\thetabm^\alphad\thetabp^\betad\varphi^{--}_{\alphad\betad}
+(\thetabp)^2 v^{(-4)}\\
&+(\thetabm)^2\thetab^+_\alphad \tau^{-\;\alphad}
+(\thetabp)^2\thetab^-_\alphad \tau^{(-3)\;\alphad}
+(\thetabp)^2(\thetabm)^2\tau^{--}\,.
\end{align}
It can be shown \cite{Ferrara:2004zv} that only $\Ac$, the coefficient of $(\thetabm)^2$ in
\eqref{curvaturaw}, contributes to the action
\begin{equation}\label{actionAComp}
    S=\frac14\int d^4x\,d^4\theta\,du\,\Ac^2.
\end{equation} 
This coefficient can be obtained from the flatness equation
\begin{equation}\label{flateqVmm}
\Dpp V^{--}-\Dmm\V+\left[ \V\,,\, V^{--}\right]_\star = 0\,.
\end{equation}
where $\V$ is the harmonic superfield which carries the $\Nc=(1,1)$
vector multiplet. In chiral coordinates we have   
\begin{equation}
\V =
v^{++} +\bar\theta^{+}_{\alphad}v^{+\alphad} +(\thetabp)^2 v\, ,
\end{equation}
\begin{subequations}
\begin{align}
\label{vpp}v^{++} = &(\theta^+)^2 \phib\, ,\\[3mm]
\label{vp}v^{+\alphad} =& 2\theta^{+\alpha}A_{\alpha}^{\alphad} +
4(\theta^+)^2\Psib^{-\alphad}
-2\iim (\theta^+)^2\theta^{-\alpha}\partial_\alpha^{\alphad}\,\phib\,,\\[3mm]
\label{v}v = &\phi + 4\theta^+\Psi^- + 3(\theta^+)^2 D^{--}
-\iim (\thetap\thetam)\partial^{\alpha\alphad}A_{\alpha\alphad}
+ \theta^{-\alpha}\theta^{+\beta}\,F_{\alpha\beta} \nonumber\\ 
&\quad-(\thetap)^2(\thetam)^2\square\phib+4\iim\,
(\thetap)^2\theta^{-\alpha}\partial_{\alpha\alphad}\Psib^{-\alphad}\,.
\end{align}
\end{subequations}
The components of \eqref{flateqVmm} relevant to determine $\Ac$ are
\begin{subequations}
\begin{align}
&\nablapp\Ac=0\,,\\[3mm]
&\nablapp v^{-\;\alphad}-v^{+\;\alphad}=0,\\[2mm]
&\nablapp\varphi^{--}+2(\Ac-v)
+\frac12\left\lbrace v^{+\;\alphad},v^-_\alphad \right\rbrace_\star =0.
\end{align}
\label{nalapps}
\end{subequations}
where
\begin{equation}
\label{nablapp}
\nablapp={\Dpp} +\left[\vpp,\right]_\star
\end{equation}
and $v^{++}$, $v^+_\alphad$ and $v$ are defined in \eqref{vpp},
\eqref{vp}, \eqref{v} respectively. The Q-deformed commutator in
\eqref{nablapp}, for a general chiral superfield $\Phi(x_L,
\theta^\pm_\alpha)\,$ (irrespective of the Grassmann parity of the
latter), reads
\begin{equation}
\left[\vpp,\Phi\right]_\star= -2\partial_{+\alpha}\vpp\partial_{+\beta}\Phi\,
b^{++}c^{\alpha\beta}-2\partial_{+\alpha}\vpp\partial_{-\beta}\Phi\,
b^{+-}c^{\alpha\beta}.
\end{equation}
Then for the product ansatz \eqref{hatcdecomp}, $\nablapp\Phi$ becomes
\begin{equation}
\nablapp\Phi=\left[\ppp -\left(\ve^{\alpha\beta}
+4\phib\bpm c^{\alpha\beta}\right)
\theta^+_\alpha\partial_{-\beta}
-4\phib\bpp\,c^{\alpha\beta}
\theta^+_\alpha\partial_{+\beta}\right]\Phi\,.
\end{equation}

It remains to solve the coupled system of equations \eqref{nalapps},
plug the solutions into \eqref{actionAComp} and integrate in the
harmonic variables using the list of integrals in Appendix B of
\cite{DeCastro:2005vb}. Due to the complexity of this calculation, it
was performed with help of a symbolic algebra computer package,
resulting into a very lengthy action in components which for our further
analysis is not necessary to present here. In fact, to show what happens
for particular values of $b^{ij}$ is much more illustrative.

\subsection{$\Nc=(1,1/2)$ supersymmetry action in components} 
Here, we consider the situation with $b^2=0$, i.e.  for those components
of $b^{ij}$ which are solutions of equation $\det{(b_{ij})}=
b_{11}b_{22}-(b_{12})^2= 0$. The action in components is
\begin{equation}
\begin{aligned}\label{Actionfullb2}
S=\int d^4x_L \Bigl[ &-\frac12\phi\square\phib -\frac1{16}
F^{\alpha\beta}F_{\alpha\beta} +\frac14 D^2+\iim
\Psi^{k\alpha}\partial_{\alpha\alphad}\Psib^{\alphad}_k+\iim
b_{ij}D^{ij}\, c^{\alpha\beta}
\partial_{(\alpha\alphad}\phib A_{\beta)}^\alphad \\[3mm]
&+\frac12\phib b_{ij}D^{ij}  c^{\alpha\beta} F_{\alpha\beta}
+\frac{4 \iim}3 b_{ij}
c^{\alpha\beta}A_{\alpha\alphad}\Psib^{i\alphad}\partial_{\beta\betad}
\Psib^{j \betad}-4\iim b_{ij}c^{\alpha}_{\beta}\Psi^{i\beta}\partial_{\alpha\alphad}\phib\Psib^{j\alphad}\\[3mm]
&-\frac43\iim b_{ij}c_{\alpha}^{\beta}\phi \Psi^{i\alpha}\partial_{\beta\alphad}\Psib^{j\alphad}+c^{\alpha\beta}F_{\alpha\beta}b_{ij}\Psib^i_{\alphad}\Psib^{j\alphad}
-4\,c^2(b_{ij}\Psib^i_{\alphad}\Psib^{j\alphad})^2 \\[3mm]
&-\frac{32}9\phib
c^2 b_{ij}D^{ij}b_{kl}\Psib^k_\alphad\Psib^{l\alphad}\Bigr].
\end{aligned}
\end{equation}
First of all, we remark that this is an \emph{exact} result for which
$3/4$ of the original supersymmetries are preserved. The main feature of
\eqref{Actionfullb2} is that we can decouple the interaction between the
scalar field $\phib$ and the gauge field and still have a deformed
action, contrary to what happens in the singlet case where decoupling
the mentioned interaction destroys the deformation
\cite{Ferrara:2004zv}. Observe also that even in this case, second order
terms in the deformation parameters appear. From the corresponding gauge
variations we directly propose the minimal Seiberg-Witten like map which
take us back to the standard form of the gauge transformations.  In
\cite{DeCastro:2005vb} we obtained the full set of exact variations,
they are 
\begin{equation}
\begin{aligned}\label{gtoriginal}
\delta\,\phib= &0\,, \qquad \delta\Psib^k_{\alphad}=0\,, \qquad
\delta\,A_{\alpha\alphad}= X\coth X \partial_{\alpha\alphad}a\, \qquad \delta D_{ij}=2\iim
b_{ij}c^{\alpha\beta}\partial_{\alpha\alphad}\phib\,
\partial^{\alphad}_\beta a\,,\\[3mm]
\delta\,\phi= &2\sqrt{c^2\,b^2}\left( \frac{1-X\coth X}{X}\right)
A^{\alpha\alphad}\partial_{\alpha\alphad}a\, ,\\[3mm]
 \delta \Psi^i_\alpha =& \Biggl\lbrace
\left[\frac{4X^2(X\coth X-1)}{X^2+\sinh^2X-X\sinh2X}
\right] b^{ij}c_{\alpha\beta}\\
&-\sqrt{c^2b^2}\,
\left[\frac{4X\cosh^2X-2X^2(\coth X+X)-\sinh2X }
{X^2+\sinh^2X-X\sinh2X}\right]\ve^{ij}\ve_{\alpha\beta}
\Biggr\rbrace \Psib_{j\alphad}\,\partial^{\beta\alphad}a\,.
\end{aligned}
\end{equation}
where $X=2\bar\phi\sqrt{c^{\alpha\beta}c_{\alpha\beta}\,b^{ik}b_{ik}}$.
Imposing $b^2=0$ we have
\begin{align}
&\delta A_{\alpha\alphad}=
\partial_{\alpha\alphad}a\,,
\qquad\delta\phi=0\,, \qquad 
\delta\Psi^i_{\beta}=-\frac43 b^{ij}c^{\alpha}_{\beta}
\Psib^{\alphad}_j\, \partial_{\alpha\alphad}a,
\qquad \delta D_{ij}=
2\iim b_{ij} c^{\alpha\beta}
\partial_{(\alpha\alphad}\phib
\partial_{\beta)}^{\alphad}a
\end{align}
Thus the Seiberg-Witten-like map becomes
\begin{equation}\label{b2}
\Psi^i_{\beta}=\widetilde{\Psi}^i_{\beta}
-\frac43b^{ij} c^{\alpha}_{\beta}\Psib^{\alphad}_j\,
{A}_{\alpha\alphad}, \qquad
D_{ij}=\widetilde{D}_{ij}
-2\iim b_{ij}c^{\alpha\beta}
{A}_{(\alpha\alphad}
\partial_{\beta)}^\alphad\phib\,.
\end{equation}
Moreover, we can further redefine $\widetilde{\Psi}^{k\alpha}$ and $\widetilde{D}^{ij}$ 
\begin{align}
\widetilde{\Psi}^{k\beta}&=\psi^{k\beta}-\frac43\iim b^{k}_{i}c_{\alpha}^{\beta}\phi \widetilde{\Psi}^{i\alpha}\\[3mm]
\widetilde{D}^{ij}&=d^{ij}-\phib\, b^{ij}c^{\alpha\beta}F_{\alpha\beta}+\frac{64}{9}\phib c^2b^{ij}b_{kl}\Psib^k_\alphad \Psib^{l\alphad}
\end{align}
to finally obtain the simple expression 
\begin{equation}
\begin{aligned}
S=\int d^4x_L \Bigl[ &-\frac12\phi\square\phib -\frac1{16}
F^{\alpha\beta}F_{\alpha\beta} +\frac14 d^2+\iim
{\psi}^{k\alpha}\partial_{\alpha\alphad}\Psib^{\alphad}_k
-4\iim b_{ij}c^{\alpha}_{\beta}\psi^{i\beta}\partial_{\alpha\alphad}\phib\Psib^{j\alphad}\\[3mm]
&+c^{\alpha\beta}F_{\alpha\beta}b_{ij}\Psib^i_{\alphad}\Psib^{j\alphad}
-4\,c^2(b_{ij}\Psib^i_{\alphad}\Psib^{j\alphad})^2 \Bigr].
\end{aligned}
\end{equation}
The last three terms are not removable under field redefinitions,
meaning we are in presence of an interacting theory. Particularly the
last two terms are of the same kind as those found in
\cite{Berkovits:2003kj}, where authors construct a deformed extension
of the low energy D3-brane super Yang-Mills action. Besides, it is very
remarkable the occurrence of an additional Yukawa-like interaction
potential. This result is also comparable with the first order action
found in \cite{Araki:2005pc}, where the authors brought up the
question, whether the exact action has higher order terms or not.  It is
clear that at least for the product ansatz \eqref{hatcdecomp} we are
able to give an answer: though we already hid almost all second order
terms appearing in the action, the last term $4\,c^2(b_{ij}
\Psib^i_{\alphad} \Psib^{j\alphad})^2$ seems to be irremovable.

\subsection{$\Nc=(1,1/2)\rightarrow \Nc=(1,0)$ supersymmetry breaking}
Turning on $b^2\neq 0$ contributions and applying the corresponding
Seiberg-Witten map, we obtain the following action up to first order in
$b^2$
\begin{equation}
\begin{aligned}\label{Accionb2}
S=\int d^4x_L \Bigl[ &-\frac12\phi\square\phib -\frac1{16}
\tilde{F}^2 +\frac14 \tilde{D}^2+\iim
\tilde{\Psi}^{k\alpha}\partial_{\alpha\alphad}\Psib^{\alphad}_k
-4\iim b_{ij}c^{\alpha}_{\beta}\tilde{\Psi}^{\beta}\partial_{\alpha\betad}\phib\Psib^{j\betad}
\\[3mm]
&-\frac{ b^2 c^2 }{6} \phib^2\tilde{F}^2+
b^2c^2 \phib^2\tilde{D}^2+c^{\alpha\beta}\tilde{F}_{\alpha\beta}b_{ij}\Psib^i_{\alphad}\Psib^{j\alphad}
+4\,c^2(b_{ij}\Psib^i_{\alphad}\Psib^{j\alphad})^2\\[3mm]&+\frac\phib2 b_{ij}\tilde{D}^{ij}  c^{\alpha\beta}\tilde{F}_{\alpha\beta}
+\frac{\phib^2 b^2}{4}(c^{\alpha\beta}\tilde{F}_{\alpha\beta})^2
-2\iim \phib b^2c^2\tilde{\Psi}^{i\alpha}\partial_{\gamma\betad}\phib\Psib^{\betad}_i\\[3mm]
& -\frac{32}9\phib
c^2 b_{ij}\tilde{D}^{ij}b_{kl}\Psib^k_\alphad\Psib^{l\alphad}+\Oh{(b^3)}\Bigr].
\end{aligned}
\end{equation}
The most important feature of this action is the non trivial interaction
term 
\begin{equation}
\frac{b^2\,c^2}6 \phib^2 \widetilde{F}^{\alpha\beta}\widetilde{F}_{\alpha\beta}
\end{equation}
These kind of interactions appearing here and in
\cite{Ferrara:2004zv, DeCastro:2005vb} can not be disentangled by a
redefinition of the fields.  In order to give an interpretation of
parameter $b^{ij}$ one can for example consider the limit
\begin{equation}\label{b22}
b_{11}=1, \qquad b_{12}=0, \qquad b_{22}\ll 1.
\end{equation}
Action  \eqref{Accionb2} can be interpreted as the weak coupling limit
of an interacting theory for $\phib$ and the gauge field, with $b_{22}$
as the coupling parameter. Another interpretation of selection
\eqref{b22} (see \cite{DeCastro:2005vb}) comes from taking
$\theta^1_\alpha$ as the left Graßmann coordinate of some ${\cal
N}=(1/2, 1/2)$ subspace of ${\cal N} = (1,1)$ superspace, i.e.
$\theta_1^\alpha \equiv \theta^\alpha$. Assuming the pseudoconjugation
for all involved quantities as in \cite{Ivanov:2003te}, and selecting
the relevant broken automorphism U(1) and O(1,1) symmetries of
${\cal N} = (1,1)$ superalgebra in such a way that $b_{ik} \equiv
(b_{11}, b_{22}, b_{12}) = (1, b_{22}, 0)\,$, the deformation operator
\eqref{Poisson} for the choice \eqref{hatcdecomp} and \mbox{$I=0$} is reduced
to $P = - \overleftarrow{\partial_\alpha} c^{\alpha\beta}
\overrightarrow{\partial_\beta} -b_{22}
\overleftarrow{\partial^2_\alpha} c^{\alpha\beta}
\overrightarrow{\partial^2_\beta}\,$. In other words, it can be expressed as
a sum of the mutually commuting chiral Poisson operators on two
different ${\cal N}=(1/2, 1/2)$ subspaces of ${\cal N} = (1,1)$
superspace, with $b_{22}$ being the ``ratio'' of two Seiberg deformation
matrices.  When $b_{22} = 0$, we fall back into the case with only one
${\cal N}=(0, 1/2)$ supersymmetry broken.  For $b_{22} \neq 0$, both
${\cal N}=(0, 1/2)$ supersymmetries are broken. The parameter $b_{22}$
measures the breakdown of the second ${\cal N}=(0, 1/2)$ supersymmetry
which is implicit in the ${\cal N}=(1/2, 1/2)$ superfield formulation
based on the  superspace $(x^m, \theta^\alpha,
\bar\theta^{\dot\alpha})\,$.  Recall that within the standard complex
conjugation the reduction to ${\cal N}=(1/2, 1/2)$ superspace makes no
sense since the latter is not closed under such conjugation
\cite{Ivanov:2003te}.

\section{Non-singlet unbroken supersymmetry transformations}
In \cite{DeCastro:2005vb} we presented detailed procedures involved in
the calculation of supersymmetry transformations and we gave a
subalgebra as an example. Here, we give the corresponding set of 
transformations to each case presented in the former section. We start
by discussing the unbroken $\Nc=(1,1/2)$ supersymmetry transformations
corresponding to the action \eqref{Actionfullb2}. We recall that this
transformations were already calculated in \cite{Araki:2005pc} by
choosing the particular matrix $\hat{C}_{11}^{\alpha\beta}$ and they are
in fact equivalent to our results when
$\hat{C}_{11}^{\alpha\beta}=c^{\alpha\beta}b_{11}$. Nevertheless, as we
pointed out before, once we have done this factorization it is
equivalent to choose any solution of $\det{(b_{ij})}=0$ recovering the
manifest R-invariant symmetry of the expressions, a worthy reason to
show the $\Nc=(1,0)$ sector as an example
\begin{align}
\delta\phib&=0, \\[3mm]
\delta A_{\alpha\alphad}&=\left[2\ve_{\alpha\beta}\ve_{ij} 
+8\phib\,c_{\alpha\beta}b_{ij}\right]
\epsilon^{i\beta}\Psib^j_{\alphad},
\end{align}
\begin{align}
\delta\Psib^i_\alphad&=
-\iim\left[\ve^{\alpha\beta}\ve^{ij}-\,4\phib\,c^{\alpha\beta}b^{ij}\right]
\epsilon_{j\beta}\partial_{\alpha\alphad}\phib,
\end{align}
\begin{align}
\delta\phi&=\Psi^i_\alpha\epsilon^j_\beta\left[ 2\ve^{\alpha\beta}\ve_{ij}
+\frac{16}3\phib c^{\alpha\beta}b_{ij}\right]
+A^\alphad_\alpha\Psib^i_\alphad\epsilon^j_\beta\left[ 
	\frac{40}3c^{\alpha\beta}b_{ij}\right]
\end{align}
\begin{align}
\delta D^{ij}&=
2\iim\partial^{\alpha\alphad}\left[
\epsilon^{(i}_{\alpha}\Psib^{j)}_{\alphad}
+4\iim \phi \epsilon^{((k\beta}\Psib^{i)}_{\alphad}c_{\alpha\beta}b^{j)}_k
-4\iim \phib^2c^2b^{ij}\epsilon^k_\alpha\Psib^l_\alphad b_{kl}\right]
\end{align}
\begin{align}
	\delta\Psi^{i\alpha}=&\biggl(
	-D^{ij}\epsilon^{\alpha\beta}
	+\biggl\{\frac12F^{\alpha\beta}
	+\frac83\left[ (b\cdot\Psib\Psib)-\phib (b\cdot D) \right]
	c^{\alpha\beta}\biggr\}\varepsilon^{ij}\nonumber\\
	&+\biggl\{\frac23\Fc^{\alpha\beta}
	+\frac13\biggl[\frac{40}3c^2\phib(b\cdot\Psib\Psib)
	-\frac{28}3c^2\phib^2(b\cdot D)
	+\frac{\sqrt{2c^2}}2\iim (A\cdot\partial\phib)
	-2\iim\phib(c\cdot G)\biggr]\varepsilon^{\alpha\beta}\nonumber\\
	&-\frac23\iim\left[ 2(A\cdot\partial\phib)
	+\phib(\partial \cdot A) \right]c^{\alpha\beta}\biggr\}b^{ij}\biggr)
	\epsilon_{j\beta}
\end{align}
For the $\Nc=(1,1/2)$ supersymmetry, we could also calculate the
$\Nc=(0,1/2)$ unbroken sector generated by $\bar{Q}_{1\alphad}$ which
would be absolutely equivalent to the exact result presented in
\cite{Araki:2005pc}. Finally we display the full unbroken $\Nc=(1,0)$
transformation laws which leaves \eqref{Accionb2} invariant. They are
\begin{subequations}
\begin{align}
\delta\phib&=0, \\[3mm]
\delta A_{\alpha\alphad}&=\left[2
\left(1+\frac43b^2c^2\phib^2\right)\ve_{\alpha\beta}\ve_{ij} 
+8\phib\,c_{\alpha\beta}b_{ij}\right]
\epsilon^{i\beta}\Psib^j_{\alphad}+\Oh(b^3),\\[3mm]
\delta\Psib^i_\alphad&=
-\iim\left[(1+4b^2c^2\phib^2)\ve^{\alpha\beta}\ve^{ij}-\,4\phib\,c^{\alpha\beta}b^{ij}\right]
\epsilon_{j\beta}\partial_{\alpha\alphad}\phib+\Oh(b^3),\\[3mm]
\delta\phi&=\Psi^i_\alpha\epsilon^j_\beta\left[ 2\ve^{\alpha\beta}\ve_{ij}
+\frac{16}3\phib c^{\alpha\beta}b_{ij}\right]
+A^\alphad_\alpha\Psib^i_\alphad\epsilon^j_\beta\left[ 
	\frac{40}3c^{\alpha\beta}b_{ij}\right]+\Oh(b^3),\\[3mm]
\delta D^{ij}&=
2\iim\partial^{\alpha\alphad}\left[\left(1+ \frac13b^2c^2\phib^2\right)
\epsilon^{(i}_{\alpha}\Psib^{j)}_{\alphad}
+4\iim \phi \epsilon^{((k\beta}\Psib^{i)}_{\alphad}c_{\alpha\beta}b^{j)}_k
-4\iim \phib^2c^2b^{ij}\epsilon^k_\alpha\Psib^l_\alphad b_{kl}\right]+\Oh(b^3),
\end{align}
\begin{align}
	\delta\Psi^{i\alpha}=&\biggl(
	-D^{ij}\epsilon^{\alpha\beta}
	+\biggl\{\left( \frac12+\frac{10}9b^2c^2\phib^2\right)F^{\alpha\beta}
	-2\iim b^2c^2\phib G^{\alpha\beta}\nonumber\\[3mm]
	&+\frac29\iim b^2c^2\phib^2\left[ 8(A\cdot\partial\phib)
	+\phib(\partial \cdot A)\right]\varepsilon^{\alpha\beta}
	+\frac23\biggl[4(b\cdot\Psib\Psib)-4\phib (b\cdot D)\nonumber\\
	&-\frac{\sqrt{2c^2}}3\iim b^2\phib(A\cdot\partial\phib)
	-\frac53b^2\phib^2(c\cdot F)-\frac23\iim b^2\phib^2(c\cdot G)
	\biggr]c^{\alpha\beta}\biggr\}\varepsilon^{ij}\nonumber\\
	&+\biggl\{\frac23\Fc^{\alpha\beta}
	+\frac13\biggl[\frac{40}3c^2\phib(b\cdot\Psib\Psib)
	-\frac{28}3c^2\phib^2(b\cdot D)
	+\frac{\sqrt{2c^2}}2\iim (A\cdot\partial\phib)
	-2\iim\phib(c\cdot G)\biggr]\varepsilon^{\alpha\beta}\nonumber\\
	&-\frac23\iim\left[ 2(A\cdot\partial\phib)
	+\phib(\partial \cdot A) \right]c^{\alpha\beta}\biggr\}b^{ij}\biggr)
	\epsilon_{j\beta}+\Oh(b^3)
\end{align}
\end{subequations}
where we have defined the following shorthands
\begin{equation}
	G_{\alpha\beta}\equiv A_{(\alpha\alphad}\partial^\alphad_{\beta)},\qquad
	\Fc^{\alpha\beta}\equiv c^{(\alpha\gamma}F_\gamma^{\beta)},\qquad
	(b\cdot\Psib\Psib)\equiv b_{ij}\Psib_\alphad^i\Psib^{i\alphad}
\end{equation}
We would like to comment that these expressions were not calculated
using series expansions on the deformation parameters. Implementing
algorithms given in \cite{DeCastro:2005vb} and using a computer program,
we actually obtained the corresponding extremely lengthy exact results
and took the appropriate limit afterwards.    

\section{Conclusions}
We have studied non-singlet Q-deformations of $\Nc=(1,1)$ gauge theories
in harmonic superspace in four Euclidean dimensions, using the
decomposition matrix $\hat{C}^{\alpha\beta}_{ij}=c^{\alpha\beta}b_{ij}$
\eqref{hatcdecomp} which preserves  space-time and R-symmetry group
U(1)$_{\text{L}}\times$ SU(2)$_{\text{R}}\times$ U(1).
Imposing the condition $b^2=0$ we built the exact expression of the
$\Nc=(1,1/2)$ action.  This Lagrangian is characterized by the presence
of interaction terms comparable with the $\Nc=1+\frac12$  deformed low
energy action of a D3-brane constructed in \cite{Berkovits:2003kj}.  It
is also worth notice that the interaction potential found has a
Yukawa-like term.  It is notable
that there are second order terms in the deformation parameters which
can not be removed by redefinition of the fields.  It is also remarkable 
that despite the complete removal of the interaction between the
scalar field $\phib$ and the gauge field, we still have a deformed
action, contrary to what happens in the singlet case where decoupling
the mentioned interaction implies the complete disappearing of
deformation \cite{Ferrara:2004zv}. We can say that this is an exclusive
feature of non-singlet deformations. 

Additionally we study the behavior of the action upon restoring the
degrees of freedom in $b^{ij}$, by analyzing the structure of the first
terms with non trivial $b^2$. We recall that this action has
$\Nc=(1,0)$, thus we conclude that we have broken the supersymmetry by
turning on the $b^2\neq0$ parameters. The most remarkable  feature of
this result is the non trivial interaction term
\begin{equation} 
\frac{b^2\,c^2}6 \phib^2
\widetilde{F}^{\alpha\beta}\widetilde{F}_{\alpha\beta} 
\end{equation}
This striking interaction, which in general characterizes Q-deformations
of $\Nc=(1,1)$ gauge multiplet (see for example \cite{Ferrara:2004zv,
DeCastro:2005vb}), is not possible to disentangle via redefinition of the
fields.  In general terms we can interpret components of $b^{ij}$ as
supersymmtry breaking tuning parameters. Turning on some of $b^{ij}$
degrees of freedom we expose the non trivial interactions between
$\phib$ and the gauge field, allowing its interpretation as coupling
constants.

It would be interesting to study the renormalizability
properties of these actions, and to find their non Abelian extensions,
as well as possible instanton solutions emerging from these theories.
Another attractive topic is to study non-singlet Q-deformed
Hypermultiplets with $b^2=0$.

\section*{Acknowledgments}
We are very grateful to Evgenyi Ivanov and Olaf Lechtenfeld for valuable
discussions, Esteban Isasi for technical support and to Boris Zupnik for
interesting email correspondence. The work of L.Q. was partially
supported by the joint DAAD-FGMA scholarship A/02/22823.  A. De C.
acknowledge the support from the \emph{Graduate College in Quantum Field
Theory, Gravitation, Statistical physics and Quantum optics, Hannover
University}.



\begin{thebibliography}{999}



%
%

\bibitem{Seiberg:2003yz}
N.~Seiberg,
JHEP {\bf 0306} (2003) 010
[arXiv:hep-th/0305248].

\bibitem{Berkovits:2003kj}
  N.~Berkovits and N.~Seiberg,
  JHEP {\bf 0307} (2003) 010
  [arXiv:hep-th/0306226].

\bibitem{deBoer:2003dn}
J.~de Boer, P.~A.~Grassi and P.~van Nieuwenhuizen,
Phys.\ Lett.\ {\bf B 574} (2003) 98
[arXiv:hep-th/0302078].

\bibitem{Ooguri:2003qp} H.~Ooguri and C.~Vafa,
Adv.\ Theor.\ Math.\ Phys.\  {\bf 7} (2003) 53
[arXiv:hep-th/0302109]; Adv.\ Theor.\ Math.\ Phys.\ {\bf 7} (2004) 405 [arXiv:hep-th/0303063].

\bibitem{Ferrara:2004zv}
S.~Ferrara, E.~Ivanov, O.~Lechtenfeld, E.~Sokatchev and B.~Zupnik,
Nucl.\ Phys.\ {\bf B 704} (2005) 154
[arXiv:hep-th/0405049].



%
%

\bibitem{Grisaru:2005we}
  M.~T.~Grisaru, S.~Penati and A.~Romagnoni,
  JHEP {\bf 0602} (2006) 043
  [arXiv:hep-th/0510175].

\bibitem{Britto:2003aj}
  R.~Britto, B.~Feng and S.~J.~Rey,
  JHEP {\bf 0308} (2003) 001
  [arXiv:hep-th/0307091].

\bibitem{Buchbinder:2005sh}
  I.~L.~Buchbinder, E.~A.~Ivanov, O.~Lechtenfeld, I.~B.~Samsonov and B.~M.~Zupnik,
  arXiv:hep-th/0511234.


%
%

\bibitem{Ferrara:2003xy}
S.~Ferrara, M.A.~Lledo and O.~Macia,
JHEP {\bf 0309} (2003) 068
[arXiv:hep-th/0307039].

\bibitem{Klemm:2001yu}
D.~Klemm, S.~Penati and L.~Tamassia, Class.\ Quant.\ Grav.\ {\bf 20}
(2003) 2905 [arXiv:hep-th/0104190].


%
%

\bibitem{Ivanov:2004qz}
  E.~A.~Ivanov and B.~M.~Zupnik,
  Theor.\ Math.\ Phys.\  {\bf 142} (2005) 197
  [Teor.\ Mat.\ Fiz.\  {\bf 142} (2005) 235]
  [arXiv:hep-th/0405185].

\bibitem{Ivanov:2003te}
E.~Ivanov, O.~Lechtenfeld and B.~Zupnik,
JHEP {\bf 0402} (2004) 012
[arXiv:hep-th/0308012].


%
%

\bibitem{Galperin:2001uw}
A.S.~Galperin, E.A.~Ivanov, V.I.~Ogievetsky and E.S.~Sokatchev,
{\it{Harmonic Superspace}}, Cambridge University Press 2001, 306 p.


%
%

\bibitem{Ferrara1}
S.~Ferrara and M.A.~Lledo, JHEP {\bf 0005} (2000)
008 [arXiv:hep-th/0002084].


%
%

\bibitem{DeCastro:2005vb}
  A.~De Castro, E.~Ivanov, O.~Lechtenfeld and L.~Quevedo,
  arXiv:hep-th/0510013.


%
%

\bibitem{Ketov:2004yt}
S.V.~Ketov and S.~Sasaki, Phys.\ Lett.\ {\bf B 595} (2004) 530 [ArXiv:hep-th/0404119];
Phys.\ Lett.\ {\bf B 597} (2004) 105 [ArXiv:hep-th/0405278];
Int.\ J.\ Mod.\ Phys.\ {\bf A 20} (2005) 4021
[arXiv:hep-th/0407211].


%
%

\bibitem{Ivanov:2004hc}
E.~Ivanov, O.~Lechtenfeld and B.~Zupnik, Nucl.\ Phys.\ {\bf B 707} (2005) 69
[arXiv:hep-th/0408146].

\bibitem{Araki:2005pc}
T.~Araki, K.~Ito and A.~Ohtsuka, JHEP 0505 (2005) 074
[arXiv:hep-th/0503224].


%
%

\bibitem{Zupnik:2006qa}
B.M.~Zupnik, 
[arXiv:hep-th/0602034].


%
%

\bibitem{Araki:2004de}
T.~Araki and K.~Ito,
Phys.\ Lett.\ B {\bf 595} (2004) 513
[arXiv:hep-th/0404250].

\bibitem{Araki:2004cv}
T.~Araki, K.~Ito and A.~Ohtsuka, Phys.\ Lett.\ {\bf B 606} (2005) 202
[arXiv:hep-th/0410203].

\bibitem{Araki:2004mq}
T.~Araki, K.~Ito and A.~Ohtsuka,
JHEP {\bf 0401} (2004) 046
[arXiv:hep-th/0401012].




\end{thebibliography}
\end{document}